\documentclass{epl}
\usepackage{graphicx}
\usepackage{amsmath}
\usepackage{amssymb}

\title{All-electronic coherent population trapping\\
in quantum dots}
\shorttitle{Population trapping in quantum dots}
\author{B. Michaelis\inst{1} \and C. Emary\inst{2} \and C.W.J. Beenakker\inst{1}}
\institute{
\inst{1} Instituut-Lorentz, Universiteit Leiden, P.O. Box 9506, 2300 RA Leiden, The Netherlands\\
\inst{2} Department of Physics, University of California San Diego, La Jolla, California 92093--0319, USA}

\begin{document}
\maketitle

\begin{abstract}
We present a fully electronic analogue of coherent population trapping in quantum optics, based on destructive interference of single-electron tunneling between three quantum dots. A large bias voltage plays the role of the laser illumination. The trapped state is a coherent superposition of the electronic charge in two of these quantum dots, so it is destabilized as a result of decoherence by coupling to external charges. The resulting current $I$ through the device depends on the ratio of the decoherence rate $\Gamma_{\phi}$ and the tunneling rates. For $\Gamma_{\phi}\rightarrow 0$ one has simply $I=e\Gamma_{\phi}$. With increasing $\Gamma_{\phi}$ the current peaks at the inverse trapping time. The direct relation between $I$ and $\Gamma_{\phi}$ can serve as a means of measuring the coherence time of a charge qubit in a transport experiment.
\end{abstract}

Coherent population trapping is a quantum optical phenomenon in which the laser illumination of an atom drives an atomic electron into a coherent superposition of orbital states and traps it there
\cite{dark1,dark2,dark3}.
Such superpositions can be ``dark'', in that they are further decoupled from the optical fields. Brandes and Renzoni have shown
how such states can also be formed in artificial atoms (quantum dots) through the use of laser illumination \cite{brandes1,brandes2}. 
In this paper we present an all-electronic analogue, i.e. without laser illumination, of coherent population trapping in quantum dots. (For an analogy in superconducting Josephson junctions, see Ref.\ \cite{Fao03,Sie04}; for an analogy in single benzene molecules, see Ref.\ \cite{Het03}.)
We illustrate this effect by considering a system of three tunnel-coupled quantum dots and show that, under proper bias and resonance 
conditions, an electron can become trapped in a coherent superposition of states in different dots.  This state is ``dark'' in the sense that, due to the Coulomb blockade, no further electrons can pass through the dots and current flow is blocked in the absence of decoherence. 

The trapping effect provides a novel mechanism for current rectification, since the blocking is effective for one sign of the bias voltage only. This quantum mechanical mechanism is distinct from mechanisms discussed previously. In particular, the classical rectification mechanism 
of Stopa and collaborators \cite{stopa02,vidan04} traps the electron in a {\em single\/} quantum dot, rather than in a coherent superposition of spatially separated states. Experiments by Ono and collaborators \cite{ono02} on rectification in double quantum dots likewise trap an electron in a single dot. The three-dot configuration requires no Aharonov-Bohm phase to trap an electron, in contrast to the two-dot configuration of Marquardt and Bruder \cite{Mar03}. Because of the phase coherent origin of the effect discussed here, the current that leaks through the device when it is blocked provides a method by which one can determine the coherence time of a charge qubit.

\begin{figure}
\centerline{\includegraphics[width=8cm]{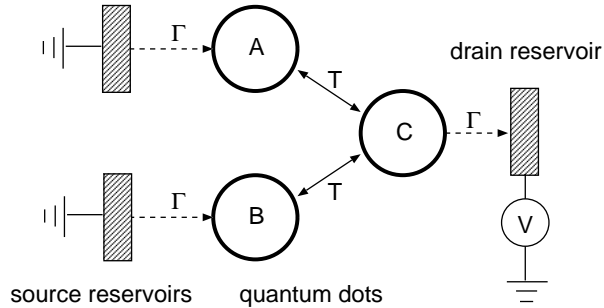}}
\caption{Schematic of the three-dot trap. The solid arrows indicate reversible transitions, described by the Hamiltonian (\ref{Hdef}). The dashed arrows indicate irreversible transitions, described by the quantum jump operator (\ref{L2def}). Destructive interference of the two reversible transitions traps an electron in a coherent superposition $|\Phi_{-}\rangle=2^{-1/2}(|A\rangle-|B\rangle)$ of the states on dots $A$ and $B$. The trapped state has vanishing amplitude on dot $C$, so that it can not decay into the right reservoir. No trapping is possible if the bias is inverted, because the trapped state would then decay into the left reservoirs.
\label{threetrapfig}
}
\end{figure}

The three-dot trap is shown schematically in Fig.\ \ref{threetrapfig}. Three quantum dots and three electron reservoirs are connected by reversible or by irreversible transitions. Reversible transitions between the quantum dots are described by the tunnel Hamiltonian
\begin{equation}
H=T|C\rangle\langle A|+T|C\rangle\langle B|+{\rm H.c.}
=2^{1/2}T|C\rangle\langle \Phi_{+}|+{\rm H.c}.
\label{Hdef}
\end{equation}
We have defined the states
\begin{equation}
|\Phi_{\pm}\rangle=2^{-1/2}(|A\rangle\pm|B\rangle).
\end{equation}
We consider the case that the energies of the single-particle levels $|A\rangle$, $|B\rangle$, $|C\rangle$ in the three dots are all the same (set at zero), so that inelastic transitions between these levels do not play a role. To minimize the number of free parameters, all tunnel rates are put equal to $T$. (The more general case of unequal tunnel rates will be considered at the end of the paper.) We assume time-reversal symmetry, hence $T$ is a real number. (Since results do not depend on the sign of $T$, we will take $T$ positive for ease of notation.) We furthermore assume that the electrostatic charging energy of the combined three-dot system is sufficiently large that the total number of electrons does not exceed one. Many-electron states are projected out and hence we may
ignore spin.

For a bias voltage $|V|\gg T/e$, and at zero temperature, the transitions from the source reservoirs into dots $A$ and $B$ and from dot $C$ into the drain reservoir are irreversible. (Because of this restriction, the rectification provided by our device does not apply to the range $|V|\lesssim T/e$ around zero bias.) The tunnel rates between dots and reservoirs are all set equal to $\Gamma$. The quantum jump operators are
\begin{equation}
L_{A}=\sqrt{\Gamma}|A\rangle\langle 0|,\;\;
L_{B}=\sqrt{\Gamma}|B\rangle\langle 0|,\;\;
L_{C}=\sqrt{\Gamma}|0\rangle\langle C|,\label{L2def}
\end{equation}
where $|0\rangle$ is the state with all three dots empty.

We study the dynamics of this device by means of the master equation approach to single-electron tunneling \cite{Naz93,sto96, gur98}, which describes not only the populations of the dot levels, but also accounts for quantum coherences between them. The master equation gives the time evolution of the three-dot density matrix $\rho(t)$ in the Lindblad form \cite{Lin76}
\begin{equation}
\frac{d\rho}{dt}=-i[H,\rho]
+\sum_{X=A,B,C}\biggl(L_{X}^{\vphantom{\dagger}}\rho L_{X}^{\dagger}-\textstyle{\frac{1}{2}}L_{X}^{\dagger}L_{X}^{\vphantom{\dagger}}\rho-\textstyle{\frac{1}{2}}\rho L_{X}^{\dagger}L_{X}^{\vphantom{\dagger}}\biggr).\label{master}
\end{equation}
(We have set $\hbar\equiv 1$.) As initial condition we take $\rho(0)=|0\rangle\langle 0|$.

We use as a basis for the density matrix the four states
\begin{equation}
|e_{1}\rangle=|\Phi_{+}\rangle,\;\; |e_{2}\rangle=|\Phi_{-}\rangle,\;\; |e_{3}\rangle=|C\rangle,\;\; |e_{4}\rangle=|0\rangle. \label{basisdef}
\end{equation}
This four-dimensional space may be reduced to a three-dimensional subspace by noting that the master equation (\ref{master}) couples only $\rho_{44}$ and $\rho_{ij}$ with $i,j\leq 3$. The matrix elements $\rho_{ij}$ with $i=4,j\neq 4$ or $j=4,i\neq 4$ remain zero. We may therefore seek a solution of the form
\begin{equation}
\rho(t)=\tilde{\rho}(t)+\left[1-{\rm Tr}\,\tilde{\rho}(t)\right]|0\rangle\langle 0|,
\end{equation}
where $\tilde{\rho}$ is restricted to the three-dimensional subspace spanned by the states $|e_{i}\rangle$ with $i\leq 3$. The evolution equation for $\tilde\rho$ reads
\begin{eqnarray}
&&\frac{d\tilde{\rho}}{dt}=M\tilde{\rho}+\tilde{\rho}M^{\dagger}+Q,\;\;\tilde{\rho}(0)=0,\label{tildemaster}\\
&&M=-\begin{pmatrix}
0&0&2^{1/2}iT\\
0&0&0\\
2^{1/2}iT&0&\Gamma/2
\end{pmatrix},\\
&&Q=\Gamma(1-{\rm Tr}\,\tilde{\rho})\begin{pmatrix}
1&0&0\\
0&1&0\\
0&0&0
\end{pmatrix}.\end{eqnarray}

All off-diagonal elements of $\tilde\rho$ vanish, except the purely imaginary $\tilde{\rho}_{13}=-\tilde{\rho}_{31}$. Four real independent variables remain, which we collect in a vector $v=({\rm Tr}\,\tilde{\rho},\tilde{\rho}_{11},\tilde{\rho}_{33},{\rm Im}\,\tilde{\rho}_{13})$ satisfying
\begin{eqnarray}
&&\frac{dv}{dt}=X\cdot(v-v_{\infty}),\;\;v(0)=0,\\
&&X=\begin{pmatrix}
-2\Gamma&0&-\Gamma&0\\
-\Gamma&0&0&-2^{3/2}T\\
0&0&-\Gamma&2^{3/2}T\\
0&2^{1/2}T&-2^{1/2}T&-\Gamma/2
\end{pmatrix},\;\;
v_{\infty}=\begin{pmatrix}
1\\0\\0\\0
\end{pmatrix}.\nonumber\\
&&
\end{eqnarray}
The solution is
\begin{equation}
v(t)=v_{\infty}-e^{Xt}v_{\infty}.
\end{equation}

All four eigenvalues $\lambda_{n}$ of $X$ have a negative real part, so $v(t)\rightarrow v_{\infty}$ for $t\rightarrow\infty$ and hence
\begin{equation}
\lim_{t\rightarrow\infty}\rho(t)=|\Phi_{-}\rangle\langle \Phi_{-}|. 
\end{equation}
This is the trapped state: it does not decay because it is an eigenstate of $H$. For large times $|v(t)-v_{\infty}|\propto e^{-\alpha t}$, with trapping rate $\alpha=\min(|{\rm Re}\,\lambda_{1}|,|{\rm Re}\,\lambda_{2}|,|{\rm Re}\,\lambda_{3}|,|{\rm Re}\,\lambda_{4}|)$. The full expression for $\alpha$ is lengthy, but the two asymptotic limits have a compact form,
\begin{equation}
\alpha=\left\{\begin{array}{cc}
4T^{2}/\Gamma&{\rm if}\;\; T\ll\Gamma,\\
\frac{1}{4}(5-\sqrt{17})\Gamma\approx 0.22\,\Gamma&{\rm if}\;\;\Gamma\ll T.
\end{array}\right.\label{alphaasymp}
\end{equation}
If the coupling of the quantum dots to the reservoirs is weaker than between themselves, then the trapping time is of order $1/\Gamma$. One might have guessed the trapping time to be of order $1/T$ in the opposite regime $T\ll\Gamma$, but this guess underestimates the correct answer, which is larger by a factor $\Gamma/T$. The fact that $\alpha\rightarrow 0$ when $\Gamma\rightarrow\infty$ can be understood as a decoherence of the inter-dot dynamics induced by a strong coupling to the electron reservoirs. 

\begin{figure}
\centerline{\includegraphics[width=8cm]{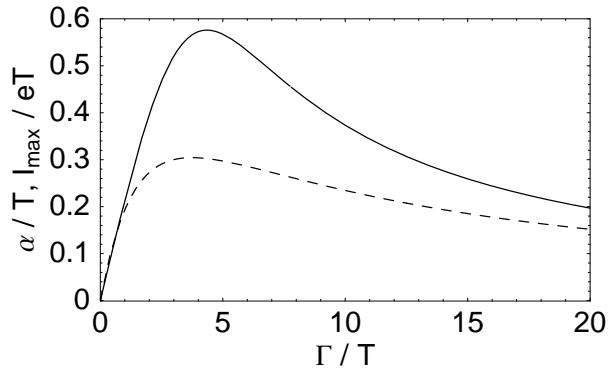}}
\caption{{\em Solid curve:} Dependence of the trapping rate $\alpha$ on the tunnel rate $\Gamma$ between quantum dots and reservoirs. Both rates are normalized by the tunnel rate $T$ between the quantum dots. The small and large-$\Gamma$ limits are given by Eq.\ (\ref{alphaasymp}). {\em Dashed curve:} Maximal steady state current $I_{\rm max}$ in the presence of decoherence, according to Eq.\ (\ref{Imax}). The two quantities $I_{\rm max}$ and $e\alpha$ differ by less than a factor of two over the whole range of tunnel rates.
\label{alphaplot}
}
\end{figure}

The full dependence of $\alpha$ on $\Gamma$ and $T$ is shown in Fig.\ \ref{alphaplot}. If $\Gamma$ is increased at constant $T$, the trapping rate has a maximum of $\alpha_{\rm max}=0.58\,T$ at $\Gamma=4.35\,T$.

The trapping effect requires the coherent superposition of spatially separated electronic states in quantum dots $A$ and $B$. Such a charge qubit is sensitive to decoherence by coupling to other charges in the environment, which effectively project the qubit on one of the three localized states $|A\rangle$, $|B\rangle$, $|C\rangle$. We model this decoherence by including into the master equation the three quantum jump operators
\begin{equation}
L_{\phi_{X}}=\Gamma_{\phi}^{1/2}|X\rangle\langle X|,\;\; X=A,B,C. 
\end{equation}
The decoherence rate $\Gamma_{\phi}$ parameterizes the strength of the charge noise and is taken to be dot independent. For a microscopic foundation of the charge noise model we refer to Ref.\ \cite{Gur97}. We also note that charge noise causes phase as well as energy relaxation.\footnote{
To calculate the energy relaxation, we decouple the three quantum dots from the electron reservoirs (setting $\Gamma\equiv 0$) and calculate $dE/dt=(d/dt){\rm Tr}\,\rho H$ from Eq.\ (\ref{master2}). One finds $dE/dt=-\Gamma_{\phi}E$, so the energy of the three-dot system relaxes to zero with rate $\Gamma_{\phi}$.}

\begin{figure}[t]
\centerline{\includegraphics[width=8cm,height=7cm]{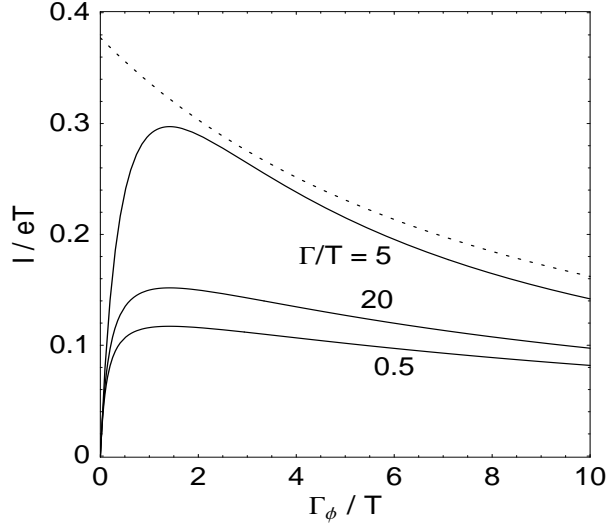}}
\caption{{\em Solid curves:} Dependence of the steady-state current $I$ on the decoherence rate $\Gamma_{\phi}$ for three values of $\Gamma/T$, calculated from Eq.\ (\ref{Isteady}). The height of the maximum depends non-monotonically on $\Gamma$, first increasing $\propto\Gamma$ and then decreasing $\propto 1/\Gamma$, according to Eq.\ (\ref{Imax}). {\em Dashed curve:} Steady-state
current if source and drain reservoir are interchanged, calculated from Eq. (\ref{Iinverted}) for $\Gamma/T=20$.
\label{Iplot}
}
\end{figure}

The master equation reads
\begin{equation}
\frac{d\rho}{dt}=-i[H,\rho]+\sum_{X=A,B,C,
\phi_{A},\phi_{B},\phi_{C}}\biggl(L_{X}^{\vphantom{\dagger}}\rho L_{X}^{\dagger}
-\textstyle{\frac{1}{2}}L_{X}^{\dagger}L_{X}^{\vphantom{\dagger}}\rho-\textstyle{\frac{1}{2}}\rho L_{X}^{\dagger}L_{X}^{\vphantom{\dagger}}\biggr).\label{master2}
\end{equation}
The steady-state current,
\begin{equation}
I=\lim_{t\rightarrow\infty}e\Gamma\langle C|\rho(t)|C\rangle,
\end{equation}
is obtained by solving Eq.\ (\ref{master2}) with the left-hand-side set to zero. We find
\begin{eqnarray}
I&=&\frac{4e\Gamma T^{2}}{\Gamma^{2}+14T^{2}+2\Gamma\Gamma_{\phi}(1+2T^{2}/\Gamma_{\phi}^{2})}\nonumber\\
&\rightarrow&\left\{\begin{array}{ll}
e\Gamma_{\phi}&{\rm if}\;\;\Gamma_{\phi}\ll\Gamma,T,\\
2eT^{2}/\Gamma_{\phi}&{\rm if}\;\;\Gamma_{\phi}\gg\Gamma,T.\end{array}
\right.\label{Isteady}
\end{eqnarray}
As illustrated in Fig.\ \ref{Iplot}, the current vanishes both in the limit $\Gamma_{\phi}\rightarrow 0$, because of the trapping effect, and in the limit $\Gamma_{\phi}\rightarrow\infty$, because of the quantum Zeno effect \cite{quantumzenotheory,quantumzenoexperiment}. The maximal current is reached at $\Gamma_{\phi}=2^{1/2}T$ and is equal to
\begin{eqnarray}
I_{\rm max}&=&\frac{4e\Gamma T^{2}}{\Gamma^{2}+14T^{2}+4\sqrt{2}\,\Gamma T}\nonumber\\
&\rightarrow&\left\{\begin{array}{ll}
4eT^{2}/\Gamma & {\rm if}\;\;T\ll\Gamma,\\
\frac{2}{7}e\Gamma\approx 0.29\,e\Gamma& {\rm if}\;\;\Gamma\ll T.
\end{array}
\right.\label{Imax}
\end{eqnarray}
Comparison of Eqs.\ (\ref{alphaasymp}) and (\ref{Imax}) shows that the maximal current $I_{\rm max}$ in the {\em presence\/} of decoherence is set by the trapping rate $\alpha$ in the {\em absence\/} of decoherence. For $T\ll\Gamma$ one has exactly $I_{\rm max}=e\alpha$, while for $\Gamma\ll T$ the two quantities differ by a numerical coefficient of order unity. In Fig.\ \ref{alphaplot} both $I_{\rm max}/e$ and $\alpha$ are plotted together, and are seen to differ by less than a factor of two over the whole $\Gamma,T$ range.

The trapping effect does not happen if the bias is inverted, so that the drain
reservoir becomes the source and vice versa. In that case we find for the
steady-state current the expression
\begin{eqnarray}
I&=&\frac{4e\Gamma T^{2}(2\Gamma+\Gamma_{\phi})}{\Gamma(\Gamma+\Gamma_{\phi})
(\Gamma+2\Gamma_{\phi})+4T^{2}(6\Gamma+5\Gamma_{\phi})}\nonumber\\
&\rightarrow&\left\{\begin{array}{ll}
8e\Gamma T^{2}(\Gamma^{2}+24T^{2})^{-1}&{\rm if}\;\;\Gamma_{\phi}\ll\Gamma,T,
\\
2eT^{2}/\Gamma_{\phi}&{\rm if}\;\;\Gamma_{\phi}\gg\Gamma,T.\end{array}
\right.\label{Iinverted}
\end{eqnarray}
For strong decoherence the current is the same in
both bias directions, but
for weak decoherence the current in the case of inverted bias does not drop
to zero but saturates at a finite value. The two cases are compared in Fig.\
\ref{Iplot}. We see that the appearance of a maximum current as a function of
$\Gamma_{\phi}$ is characteristic for the trapping effect.

We have for simplicity assumed that all three dots have the same tunnel rates and decoherence rates, but this assumption may be easily relaxed. Let us consider first the case that the three-dot structure still has a reflection symmetry, so that dots $A$ and $B$ are equivalent, but that dot $C$ has a different tunnel rate $\Gamma'$ into the reservoir and a different decoherence rate $\Gamma'_{\phi}$. We denote $\bar{\Gamma}_{\phi}=(\Gamma_{\phi}+\Gamma'_{\phi})/2$. The result (\ref{Isteady}) for the steady-state current generalizes to
\begin{eqnarray}
I&=&\frac{4e\Gamma' T^{2}}{\Gamma'^{2}+2T^{2}(6+\Gamma'/\Gamma)+2\Gamma'\Gamma_{\phi}(\bar{\Gamma}_{\phi}/\Gamma_{\phi}+2T^{2}/\Gamma_{\phi}^{2})}\nonumber\\
&\rightarrow&\left\{\begin{array}{ll}
e\Gamma_{\phi}&{\rm if}\;\;\Gamma_{\phi}\rightarrow 0,\\
2eT^{2}/\bar{\Gamma}_{\phi}&{\rm if}\;\;\Gamma_{\phi}\rightarrow\infty. \end{array}
\right.\label{Isteady2}
\end{eqnarray}
The steady-state current still contains the desired information on the rates of decoherence, with the regimes of weak and strong decoherence governed by $\Gamma_{\phi}$ and $\bar{\Gamma}_{\phi}$, respectively. 

In the most general case of arbitrarily different tunnel rates $T_{A},T_{B},\Gamma_{A},\Gamma_{B},\Gamma_{C}$ and decoherence rates $\Gamma_{\phi_{A}},\Gamma_{\phi_{B}},\Gamma_{\phi_{C}}$, the steady state current in the limit of weak and strong decoherence takes the form
\begin{subequations}
\label{Isteady3}
\begin{eqnarray}
I&\rightarrow&\frac{w_{0}e(\Gamma_{\phi_{A}}+\Gamma_{\phi_{B}})}{w_{A}+w_{B}}\;\;{\rm if}\;\;\Gamma_{\phi}\rightarrow 0,\label{Isteady3a}\\
I&\rightarrow&\frac{4eT_{A}T_{B}}{w_{A}\Gamma_{\phi_{A}}+w_{B}\Gamma_{\phi,B}+(w_{A}+w_{B})\Gamma_{\phi_{C}}}
\;\;{\rm if}\;\;\Gamma_{\phi}\rightarrow\infty,\nonumber\\
&&\label{Isteady3b}
\end{eqnarray}
\end{subequations}
with weight factors
\begin{equation}
w_{0}=\frac{T_{A}T_{B}}{T_{A}^{2}+T_{B}^{2}},\;\;w_{A}=\frac{\Gamma_{A}T_{B}/T_{A}}{\Gamma_{A}+\Gamma_{B}},\;\;w_{B}=\frac{\Gamma_{B}T_{A}/T_{B}}{\Gamma_{A}+\Gamma_{B}},\label{wdef}
\end{equation}
that are functions of the tunnel rates --- but independent of the decoherence rates. Notice that in this asymmetric case the trapped state $\sqrt{w_{0}T_{B}/T_{A}}|A\rangle-\sqrt{w_{0}T_{A}/T_{B}}|B\rangle$ has unequal weights on the two dots $A$ and $B$.

In conclusion, we have demonstrated how the well known concept of coherent population trapping in atoms may be transferred to a purely electronic system. A large voltage bias plays the role of the laser illumination and single-electron tunneling between quantum dots plays the role of intra-atomic transitions. Because the quantum dots are charged, the trapped electronic state is sensitive to decoherence by coupling to charges in the environment. This decoherence destabilizes the trapped state, causing a leakage current $I$ to flow through the quantum dots. We have found that the maximal $I$ in the presence of decoherence is set by the trapping rate $\alpha$, with $I_{\rm max}\approx e\alpha$ within a factor of two over the whole parameter range. For small decoherence rate $\Gamma_{\phi}$ we find $I=e\Gamma_{\phi}$, which provides a way to measure the coherence time of a charge qubit in a transport experiment. We finally note that extensions to many-electron trapping can serve as a source for the formation of entangled electron pairs \cite{Fab04}.

\acknowledgments
We have benefited from discussions with B. Trauzettel. This work was supported by the Dutch Science Foundation NWO/FOM and the US NSF project DMR 0403465.

\end{document}